\documentclass[prb,preprint,nofootinbib]{revtex4-2} 

\usepackage{amsmath}
\usepackage{amsfonts}
\usepackage{graphicx}
\usepackage[colorlinks]{hyperref}

\begin{document}

\title{A note on the planar triangles in Minkowski spacetime}


\author{Yan Cao}
\email{caoyan21@mails.ucas.ac.cn}
\affiliation{School of Astronomy and Space Sciences, University of Chinese Academy of Sciences (UCAS), Beijing 100049, China}

\date{\today}

\begin{abstract}
The geometry of 2D Minkowski spacetime $\mathbb{R}^{1,1}$ (or Minkowski plane\footnote{This is not to be confused with the 2D non-Euclidean normed space \cite{2016arXiv160206144L}, also called the Minkowski plane.}) is similar but fundamentally different from the more familiar Euclidean plane geometry \cite{2007arXiv0712.2234A,10.1119/1.3480023,Catoni2011,Boskoff2020}. This note gives an elementary discussion on some basic properties of a triangle on the Minkowski plane. In particular, we show that the theorem of Feuerbach\footnote{In 1822, Feuerbach discovered the nine-point circle and also proved that it is internally tangent to the incircle and externally tangent to the excircles of the triangle \cite{Mackay_1892}, see \cite{buba2004analogues,lgorzata2005monge,akopyan2011some,2016arXiv160206144L,giang2023some,avksentyev2023feuerbach} for some extensions of this theorem within or beyond the Euclidean plane geometry.} also holds and a use of the incenter/excenters is pointed out.
\end{abstract}

\maketitle

Taking the speed of light $c=1$, the Minkowski plane can be described by the line element $ds^2=dx^2-dt^2$ with a set of spacetime coordinate $(x,t)$, it is both flat and translationally invariant, thus has straight geodesics; physically these correspond to the worldlines of constant-velocity free-falling observers, which can be timelike ($|\Delta t/\Delta x|>1$), null ($|\Delta t/\Delta x|=1$) or spacelike ($|\Delta t/\Delta x|<1$). We define the distance between two points $(x_1,t_1)$ and $(x_2,t_2)$ (or the \textit{length} of a vector $(x_2-x_1,t_2-t_1)$ connecting them) as $\sqrt{(x_1-x_2)^2-(t_1-t_2)^2}$, with positive real or imaginary part. One can also use a polar chart $(r,\theta)$ defined by $\vec r=(x,t)=r(\sinh \theta, \cosh \theta)$ with $r,\theta\in \mathbb{R}$ if $\vec r$ is timelike, the length $|\vec r|$ of the vector $\vec r$ is then $i|r|$. Given two vectors $\vec a=(x_a,t_a)$ and $\vec b=(x_b,t_b)$, a natural definition for their inner product (from the metric) is $\vec a \cdot \vec b= x_ax_b-t_at_b$, such that $|\vec r|^2=\vec r \cdot \vec r$. Using this inner product, the angle between two timelike vectors can be defined as $\angle(\vec a, \vec b)=-\text{sgn}(\vec a \cdot \vec b)\,\text{arccosh}\left[\vec a \cdot \vec b/(|\vec a|\,|\vec b|)\right]$ (with $\text{arccosh}\,x>0$), such that for $\vec a=r_a(\sinh \theta_a, \cosh \theta_a)$ and $\vec b=r_b(\sinh \theta_b, \cosh \theta_b)$, $\angle(\vec a, \vec b)=|\theta_a-\theta_b|\,\text{sgn}(r_ar_b)$, since $\vec a\cdot\vec b<0$. The angle between a timelike vector $\vec a$ and a spacelike vector $\vec b$ can be defined as $\angle(\vec a, \vec b)=\text{arcsinh}\left[i(\vec a \cdot \vec b)/(|\vec a|\,|\vec b|)\right]$, such that for $\vec a=r_a(\sinh \theta_a, \cosh \theta_a)$ and $\vec b=r_b(\cosh \theta_b, \sinh \theta_b)$, $\angle(\vec a, \vec b)=(\theta_a-\theta_b)\,\text{sgn}(r_ar_b)$. Both the distance and the included angle are invariant under the coordinate transformation so called Lorentz boost, e.g.,
\begin{equation}
\left(\begin{matrix}t'\\x'\end{matrix}\right)
=
e^{\alpha\left(\begin{smallmatrix}0&1\\ 1&0\end{smallmatrix}\right)}
\left(\begin{matrix}t\\x\end{matrix}\right)
=\frac{\left(\begin{smallmatrix}1&\beta\\ \beta&1\end{smallmatrix}\right)}{\sqrt{1-\beta^2}}
\left(\begin{matrix}t\\x\end{matrix}\right)
=
\left(\begin{matrix}\cosh \alpha& \sinh\alpha\\  \sinh\alpha&\cosh \alpha\end{matrix}\right)
\left(\begin{matrix}r\cosh \theta\\ r\sinh\theta\end{matrix}\right)=\left(\begin{matrix}r\cosh \theta'\\ r\sinh \theta'\end{matrix}\right),
\end{equation}
(where $\beta$ is the boost's velocity and $\alpha$ is the so called rapidity) which leads to a hyperbolic rotation: $\theta'=\theta+\alpha$, while preserving the inner product.

\newpage
Two vectors are orthogonal to each other if their inner product vanishes; a timelike vector is orthogonal to a spacelike vector, while a null vector is orthogonal to itself. Given three points $\vec A=(x_A,t_A)$, $\vec B=(x_B,t_B)$, $\vec C=(x_C,t_C)$, it is straightforward to show that there is a unique point $H_A$ on the line $BC$ satisfying $(\vec H_A-\vec A)\cdot(\vec B-\vec C)=0$, which is the perpendicular feet dropped from $A$ on $BC$. The distance between $A$ and the line $BC$, defined as the length of $\vec H_A -\vec A$, is given explicitly by
\begin{equation}
|\vec H_A -\vec A|=\sqrt{-\frac{[x_B(t_C-t_A)+x_C(t_B-t_A)+x_A(t_B-t_C)]^2}{(x_B-x_C)^2-(t_B-t_C)^2}}.
\end{equation}
If the distances between a point $C'$ and both $CB$ and $CA$ are same, then $C'$ lies in the bisector of $\angle (\vec A-\vec C,\vec B-\vec C)$, this bisector can be constructed as follows. First we find a point $\vec r_*=(x_*,t_*)$ on the line $BC$ satisfying $|\vec r_*-\vec C|=|\vec A-\vec C|$, the line passing through both $\vec C$ and $(\vec A+\vec r_*)/2$ then gives the bisector, explicitly we have
\begin{equation}\label{x_*}
x_*^\pm=x_c\pm (x_B-x_C)\sqrt{\frac{|\vec A-\vec C|^2}{|\vec B -\vec C|^2}},
\quad
t_*^\pm=\frac{t_B-t_C}{x_B-x_C}(x_*^\pm-x_C)+t_C.
\end{equation}
The bisector of $\angle (\vec A-\vec C,\vec B-\vec C)$ is $\frac{t-t_C}{x-x_C}=\frac{(t_*^++t_A)/2-t_C}{(x_*^++x_A)/2-x_C}\equiv f_+(A,B,C)$, while the bisector of its supplementary angle is $\frac{t-t_C}{x-x_C}=\frac{(t_*^-+t_A)/2-t_C}{(x_*^-+x_A)/2-x_C}\equiv f_-(A,B,C)$. The incenter of triangle $\triangle ABC$ is given by the intersection point of three inner bisectors, namely
\begin{equation}
\frac{t-t_C}{x-x_C}=f_+(A,B,C),
\quad
\frac{t-t_A}{x-x_A}=f_+(B,C,A),
\quad
\frac{t-t_B}{x-x_B}=f_+(C,A,B).
\end{equation}
It is inside the triangle. As can be seen from Eq.~\eqref{x_*}, only when $|\vec A-\vec C|^2, |\vec A-\vec B|^2, |\vec B-\vec C|^2$ have the same sign does the incenter exist, that is, we can only talk about the incenter of a timelike/spacelike triangle, the all sides of which are timelike/spacelike.

\begin{figure}[hbt]
	\centering
	\includegraphics[width=0.368\textwidth]{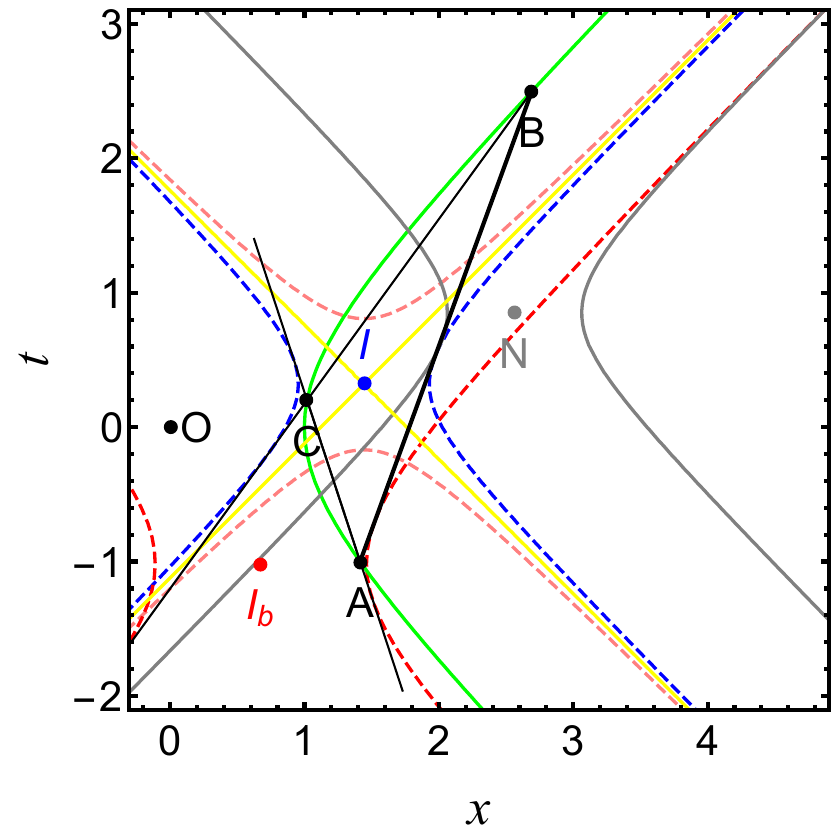}
	\qquad\qquad
	\includegraphics[width=0.356\textwidth]{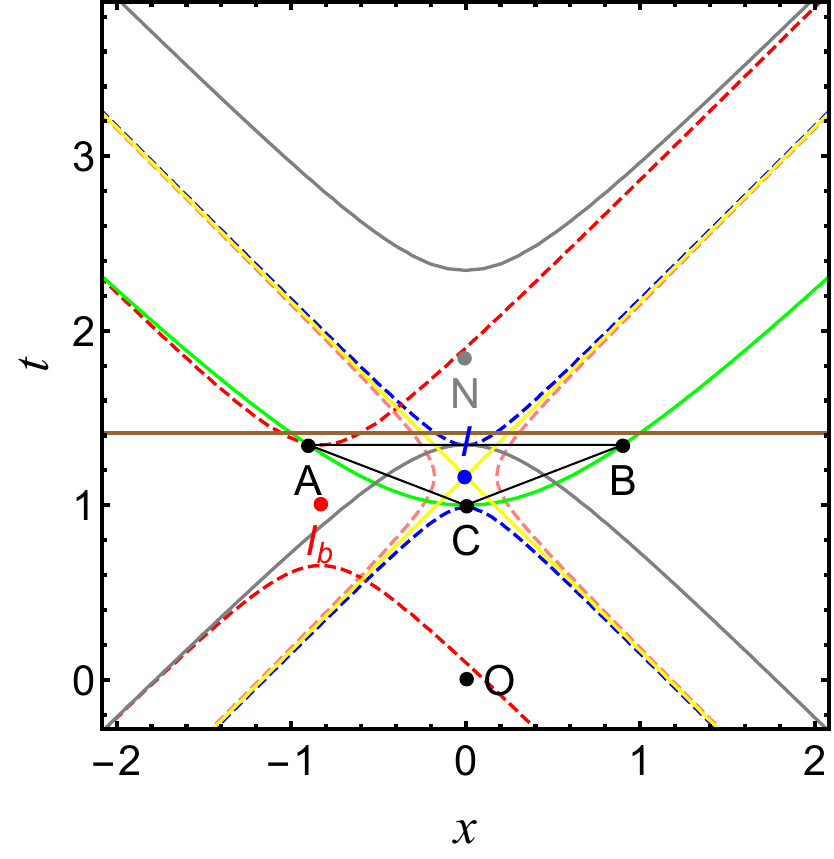}
	\caption{Minkowski triangle mapped into a Euclidean plane. Left: a timelike triangle. Right: a spacelike triangle. The red and blue dashed curves correspond to the incircle (centered at $I$) and an excircle (centered at $I_b$), respectively; the yellow lines are null trajectories passing through $I$; the gray curve is the nine-point circle (centered at $N$). Actually by switching the $x$ and $t$ coordinates we get a spacelike diagram from a timelike diagram, they are symmetrical about the line $x=t$. If $A,B,C$ are not at the same branch of the circumcircle (green curve), we get a mixed triangle.}\label{plot}
\end{figure}

The Minkowski plane can be mapped into a Euclidean plane by simply identifying $(x,t)$ with the Euclidean coordinate $(x,y)$, which is the Minkowski diagram. Under this mapping, a straight line remains straight, a Minkowski \textit{circle}\footnote{Physically, a Minkowski circle $x^2-t^2=r^2$ with real radius $r$ is associated with a Rindler observer with constant proper acceleration $a=1/r$, i.e., $\frac{d}{dt}\left(\frac{\dot x}{\sqrt{1-\dot x^2}}\right)=a$, the solution can be parameterized by the proper time $d\tau=\sqrt{1-\dot x^2}\,dt$ as $x(\tau)=x_0+a^{-1}[\cosh(a\tau)-1]$, $t(\tau)=t_0+a^{-1}\sinh (a\tau)$, and therefore $(x-x_0+a^{-1})^2-(t-t_0)^2=a^{-2}$.} (the curve with constant distance to a point), however, turns into a Euclidean equilateral hyperbola. Once we draw a Minkowski triangle $\triangle ABC$, just as the Euclidean case, there are several Minkowski circles associated with it, these include the incircle (with radius $r_\star$) centered at the incenter $I$, the excircles (with radius $r_{a,b,c}$) centered at the excenters $I_{a,b,c}$ opposite to the vertex $A,B,C$, respectively, and the circumcircle\footnote{It is easy to show that the circumcircle exists for a generic triangle without null sides.} (with radius $R$) centered at the circumcenter $O$, specifically we choose $\vec O=(0,0)$ such that $|\vec A|=|\vec B|=|\vec C|=R$. The definitions of these circles are akin to their Euclidean versions, only with the condition of tangency defined in the Minkowski sense, and the incenter/excenter exists only for a non-mixed (i.e., timelike or spacelike) triangle. We denote the sidelengths of the triangle by $a=|\vec B-\vec C|$, $b=|\vec A - \vec C|$, $c=|\vec A -\vec B|$, and the semiperimeter by $s=(a+b+c)/2$. Perhaps not surprisingly, Heron's formula $\Delta=\sqrt{s(s-a)(s-b)(s-c)}$ is still valid, where $\Delta$ is the \textit{area} of $\triangle ABC$ defined by $\Delta=\frac{1}{2}\times\text{Base}\times\text{Height}=\frac{1}{2}|\vec B-\vec C||\vec H_A-\vec A|=\frac{i}{2} |(x_B - x_A) (t_C - t_A) - (t_B - t_A) (x_C - x_A)|$. Moreover, one finds that
\begin{equation}
\vec I=\frac{a}{2s}\vec{A}+\frac{b}{2s}\vec{B}+\frac{c}{2s}\vec{C},
\quad
\vec I_c=\frac{a}{2(s-c)}\vec{A}+\frac{b}{2(s-c)}\vec{B}-\frac{c}{2(s-c)}\vec{C},
\end{equation}
with
\begin{equation}\label{radius}
r_\star=\frac{\Delta}{s}
,
\quad
r_c=\lambda_c\frac{\Delta}{s-c}
,
\quad
R=-\frac{abc}{\Delta}
,
\quad
\lambda_c=\text{sgn}(|a|+|b|-|c|).
\end{equation}
In comparison, the Euclidean result \cite{hofbauer2016simple} is $r_c=\frac{\Delta}{s-c}$ with $R=\frac{abc}{\Delta}$. The area is imaginary, for a timelike triangle, the side lengths are imaginary hence $\{R,r_\star,r_c\}$ are real. The orthocenter $H$ is given by $\vec H= \vec A+ \vec B +\vec C$, and for a non-mixed triangle turns out to be the excenter of its pedal triangle opposite to $H_\Lambda$, where $\Lambda$ is the only side with $\lambda_\Lambda=-1$; for a mixed triangle, $\lambda$ can  be positive for all three sides, in this case $\vec H$ is the incenter of its pedal triangle. Finally, the nine-point circle — a circle that passes through $H_{A,B,C}$,$\frac{\vec A+\vec B}{2}$, $\frac{\vec A+\vec C}{2}$, $\frac{\vec B+\vec C}{2}$, $\frac{\vec A+\vec H}{2}$, $\frac{\vec B+\vec H}{2}$, $\frac{\vec C+\vec H}{2}$ — exists, it is centered at $\vec N=\frac{1}{2}\vec H$ (on the Euler line) and has radius $R/2$. Following the same approach of \cite{hofbauer2016simple}, we can obtain
\begin{equation}
|\vec I -\vec N|=R/2+r_\star,
\quad
|\vec I_c -\vec N|=\sqrt{(R/2-\lambda_c r_c)^2},
\end{equation}
which gives the Feuerbach theorem on Minkowski plane. In comparison, the Euclidean result is $|\vec I -\vec N|=R/2-r_\star$, $|\vec I_c -\vec N|=R/2+r_c$. Examples of Minkowski triangle and its nine-point circle are depicted in Fig.~\ref{plot}.

The incenter and excenters of a non-mixed Minkowski triangle have a simple physical interpretation. The incircle is tangent to all three sides of the triangle (or their extensions), if we denote the points of tangency by $T_{A,B,C}$, then $|\vec T_A-\vec I|=|\vec T_B-\vec I|=|\vec T_C-\vec I|=r_\star$. Consider three ships with constant colinear velocity whose worldlines form a timelike triangle $\triangle ABC$, if a light source is placed on its incenter, the light pulse would take exactly same amount of time to reach these ships in their comoving frames, which is $\Delta t=r_\star$. One can also conceive the inverse process. The situation is similar for the three excenters.

Some of the results presented above for the triangles (2-simplices) can possibly be extended to the higher dimensional Minkowski spacetime $\mathbb{R}^{1,n}$, where the Minkowski sphere with imaginary radius is a hyperbolic space $\mathbb{H}^n$, and the sphere with real radius $1/H$ is a de Sitter spacetime $\text{dS}^n$.

\appendix
\section{Pseudo-Complex Representation of $H_A$}
Just as a point $(x,y)$ on the Euclidean plane can be represented by a complex number $z=x+iy$ such that $zz^*=x^2+y^2$, a point $(x,t)$ on the Minkowski plane can be represented by a pseudo-complex number (see for example \cite{hess2016pseudo,Catoni2011}) $z=x+\mathcal{I}t$ with $z^*=x-\mathcal{I}t$ and $\mathcal{I}^2=1$ such that $zz^*=x^2-t^2$ and $\cosh \theta +\mathcal{I}\sinh \theta=e^{\mathcal{I}\theta}$.

In this pseudo-complex (pc) representation, we can write $A=x_A+\mathcal{I} t_A$, $B=x_B+\mathcal{I} t_B$, $C=x_C+\mathcal{I} t_C$ with the circumcenter $O=0$, then for circumradius $R=\sqrt{\pm 1}$, $H_A=\frac{H\mp A^*BC}{2}$, where $H=A+B+C$. This can be proved by showing that $\text{Im}[(B-H_A)(C-H_A)^*]=\text{Re}[(B-H_A)(A-H_A)^*]=0$, where $\text{Re}$/$\text{Im}$ denotes the real/imaginary part of a pc number.


\bibliography{paper}

\begin{thebibliography}{13}%
\makeatletter
\providecommand \@ifxundefined [1]{%
 \@ifx{#1\undefined}
}%
\providecommand \@ifnum [1]{%
 \ifnum #1\expandafter \@firstoftwo
 \else \expandafter \@secondoftwo
 \fi
}%
\providecommand \@ifx [1]{%
 \ifx #1\expandafter \@firstoftwo
 \else \expandafter \@secondoftwo
 \fi
}%
\providecommand \natexlab [1]{#1}%
\providecommand \enquote  [1]{``#1''}%
\providecommand \bibnamefont  [1]{#1}%
\providecommand \bibfnamefont [1]{#1}%
\providecommand \citenamefont [1]{#1}%
\providecommand \href@noop [0]{\@secondoftwo}%
\providecommand \href [0]{\begingroup \@sanitize@url \@href}%
\providecommand \@href[1]{\@@startlink{#1}\@@href}%
\providecommand \@@href[1]{\endgroup#1\@@endlink}%
\providecommand \@sanitize@url [0]{\catcode `\\12\catcode `\$12\catcode
  `\&12\catcode `\#12\catcode `\^12\catcode `\_12\catcode `\%12\relax}%
\providecommand \@@startlink[1]{}%
\providecommand \@@endlink[0]{}%
\providecommand \url  [0]{\begingroup\@sanitize@url \@url }%
\providecommand \@url [1]{\endgroup\@href {#1}{\urlprefix }}%
\providecommand \urlprefix  [0]{URL }%
\providecommand \Eprint [0]{\href }%
\providecommand \doibase [0]{https://doi.org/}%
\providecommand \selectlanguage [0]{\@gobble}%
\providecommand \bibinfo  [0]{\@secondoftwo}%
\providecommand \bibfield  [0]{\@secondoftwo}%
\providecommand \translation [1]{[#1]}%
\providecommand \BibitemOpen [0]{}%
\providecommand \bibitemStop [0]{}%
\providecommand \bibitemNoStop [0]{.\EOS\space}%
\providecommand \EOS [0]{\spacefactor3000\relax}%
\providecommand \BibitemShut  [1]{\csname bibitem#1\endcsname}%
\let\auto@bib@innerbib\@empty
\bibitem [{\citenamefont {Aceff-Sanchez}\ and\ \citenamefont
  {Senior}(2007)}]{2007arXiv0712.2234A}%
  \BibitemOpen
  \bibfield  {author} {\bibinfo {author} {\bibfnamefont {F.}~\bibnamefont
  {Aceff-Sanchez}}\ and\ \bibinfo {author} {\bibfnamefont {L.~D.~R.}\
  \bibnamefont {Senior}},\ }\href {https://arxiv.org/abs/0712.2234} {\bibinfo
  {title} {Geometry of the conics on the minkowski plane}} (\bibinfo {year}
  {2007}),\ \Eprint {https://arxiv.org/abs/0712.2234} {arXiv:0712.2234
  [math-ph]} \BibitemShut {NoStop}%
\bibitem [{\citenamefont {Boozer}(2010)}]{10.1119/1.3480023}%
  \BibitemOpen
  \bibfield  {author} {\bibinfo {author} {\bibfnamefont {A.~D.}\ \bibnamefont
  {Boozer}},\ }\bibfield  {title} {\bibinfo {title} {{Periodic lattices in
  Minkowski space}},\ }\href {https://doi.org/10.1119/1.3480023} {\bibfield
  {journal} {\bibinfo  {journal} {American Journal of Physics}\ }\textbf
  {\bibinfo {volume} {78}},\ \bibinfo {pages} {1379} (\bibinfo {year}
  {2010})},\ \Eprint
  {https://arxiv.org/abs/https://pubs.aip.org/aapt/ajp/article-pdf/78/12/1379/12837498/1379\_1\_online.pdf}
  {https://pubs.aip.org/aapt/ajp/article-pdf/78/12/1379/12837498/1379\_1\_online.pdf}
  \BibitemShut {NoStop}%
\bibitem [{\citenamefont {Catoni}\ \emph {et~al.}(2011)\citenamefont {Catoni},
  \citenamefont {Boccaletti}, \citenamefont {Cannata}, \citenamefont {Catoni},\
  and\ \citenamefont {Zampetti}}]{Catoni2011}%
  \BibitemOpen
  \bibfield  {author} {\bibinfo {author} {\bibfnamefont {F.}~\bibnamefont
  {Catoni}}, \bibinfo {author} {\bibfnamefont {D.}~\bibnamefont {Boccaletti}},
  \bibinfo {author} {\bibfnamefont {R.}~\bibnamefont {Cannata}}, \bibinfo
  {author} {\bibfnamefont {V.}~\bibnamefont {Catoni}},\ and\ \bibinfo {author}
  {\bibfnamefont {P.}~\bibnamefont {Zampetti}},\ }\bibinfo {title}
  {Trigonometry in the hyperbolic (minkowski) plane},\ in\ \href
  {https://doi.org/10.1007/978-3-642-17977-8_4} {\emph {\bibinfo {booktitle}
  {Geometry of Minkowski Space-Time}}}\ (\bibinfo  {publisher} {Springer Berlin
  Heidelberg},\ \bibinfo {address} {Berlin, Heidelberg},\ \bibinfo {year}
  {2011})\ pp.\ \bibinfo {pages} {33--55}\BibitemShut {NoStop}%
\bibitem [{\citenamefont {Boskoff}\ and\ \citenamefont
  {Capozziello}(2020)}]{Boskoff2020}%
  \BibitemOpen
  \bibfield  {author} {\bibinfo {author} {\bibfnamefont {W.-G.}\ \bibnamefont
  {Boskoff}}\ and\ \bibinfo {author} {\bibfnamefont {S.}~\bibnamefont
  {Capozziello}},\ }\bibinfo {title} {Basic facts in euclidean and minkowski
  plane geometry},\ in\ \href {https://doi.org/10.1007/978-3-030-47894-0_2}
  {\emph {\bibinfo {booktitle} {A Mathematical Journey to Relativity: Deriving
  Special and General Relativity with Basic Mathematics}}}\ (\bibinfo
  {publisher} {Springer International Publishing},\ \bibinfo {address} {Cham},\
  \bibinfo {year} {2020})\ pp.\ \bibinfo {pages} {29--38}\BibitemShut {NoStop}%
\bibitem [{\citenamefont {Leopold}\ and\ \citenamefont
  {Martini}(2016)}]{2016arXiv160206144L}%
  \BibitemOpen
  \bibfield  {author} {\bibinfo {author} {\bibfnamefont {U.}~\bibnamefont
  {Leopold}}\ and\ \bibinfo {author} {\bibfnamefont {H.}~\bibnamefont
  {Martini}},\ }\href {https://arxiv.org/abs/1602.06144} {\bibinfo {title}
  {Monge points, euler lines, and feuerbach spheres in minkowski spaces}}
  (\bibinfo {year} {2016}),\ \Eprint {https://arxiv.org/abs/1602.06144}
  {arXiv:1602.06144 [math.MG]} \BibitemShut {NoStop}%
\bibitem [{\citenamefont {Mackay}(1892)}]{Mackay_1892}%
  \BibitemOpen
  \bibfield  {author} {\bibinfo {author} {\bibfnamefont {J.~S.}\ \bibnamefont
  {Mackay}},\ }\bibfield  {title} {\bibinfo {title} {History of the nine-point
  circle},\ }\href {https://doi.org/10.1017/S0013091500031163} {\bibfield
  {journal} {\bibinfo  {journal} {Proceedings of the Edinburgh Mathematical
  Society}\ }\textbf {\bibinfo {volume} {11}},\ \bibinfo {pages} {19–57}
  (\bibinfo {year} {1892})}\BibitemShut {NoStop}%
\bibitem [{\citenamefont {Buba-Brzozowa}(2004)}]{buba2004analogues}%
  \BibitemOpen
  \bibfield  {author} {\bibinfo {author} {\bibfnamefont {M.}~\bibnamefont
  {Buba-Brzozowa}},\ }\bibfield  {title} {\bibinfo {title} {Analogues of the
  nine-point circle for orthocentric n-simplexes},\ }\href@noop {} {\bibfield
  {journal} {\bibinfo  {journal} {Journal of Geometry}\ }\textbf {\bibinfo
  {volume} {81}},\ \bibinfo {pages} {21} (\bibinfo {year} {2004})}\BibitemShut
  {NoStop}%
\bibitem [{\citenamefont {lgorzata Buba-Brzozowa}(2005)}]{lgorzata2005monge}%
  \BibitemOpen
  \bibfield  {author} {\bibinfo {author} {\bibfnamefont {M.}~\bibnamefont
  {lgorzata Buba-Brzozowa}},\ }\bibfield  {title} {\bibinfo {title} {The monge
  point and the 3 (n+ 1) point sphere of an n-simplex},\ }\href@noop {}
  {\bibfield  {journal} {\bibinfo  {journal} {Journal for Geometry and
  Graphics}\ }\textbf {\bibinfo {volume} {9}},\ \bibinfo {pages} {31} (\bibinfo
  {year} {2005})}\BibitemShut {NoStop}%
\bibitem [{\citenamefont {Akopyan}(2011)}]{akopyan2011some}%
  \BibitemOpen
  \bibfield  {author} {\bibinfo {author} {\bibfnamefont {A.~V.}\ \bibnamefont
  {Akopyan}},\ }\href {https://arxiv.org/abs/1105.2153} {\bibinfo {title} {On
  some classical constructions extended to hyperbolic geometry}} (\bibinfo
  {year} {2011}),\ \Eprint {https://arxiv.org/abs/1105.2153} {arXiv:1105.2153
  [math.MG]} \BibitemShut {NoStop}%
\bibitem [{\citenamefont {GIANG}\ and\ \citenamefont
  {LE~VIET}(2023)}]{giang2023some}%
  \BibitemOpen
  \bibfield  {author} {\bibinfo {author} {\bibfnamefont {N.~N.}\ \bibnamefont
  {GIANG}}\ and\ \bibinfo {author} {\bibfnamefont {A.}~\bibnamefont
  {LE~VIET}},\ }\bibfield  {title} {\bibinfo {title} {Some generalizations of
  the feuerbach's theorem.},\ }\href@noop {} {\bibfield  {journal} {\bibinfo
  {journal} {Global Journal of Advance Research on Classical \& Modern
  Geometries}\ }\textbf {\bibinfo {volume} {12}} (\bibinfo {year}
  {2023})}\BibitemShut {NoStop}%
\bibitem [{\citenamefont {Avksentyev}(2023)}]{avksentyev2023feuerbach}%
  \BibitemOpen
  \bibfield  {author} {\bibinfo {author} {\bibfnamefont {E.~A.}\ \bibnamefont
  {Avksentyev}},\ }\href {https://arxiv.org/abs/2301.00731} {\bibinfo {title}
  {Feuerbach's and poncelet's theorems meet in space}} (\bibinfo {year}
  {2023}),\ \Eprint {https://arxiv.org/abs/2301.00731} {arXiv:2301.00731
  [math.DS]} \BibitemShut {NoStop}%
\bibitem [{\citenamefont {Hofbauer}(2016)}]{hofbauer2016simple}%
  \BibitemOpen
  \bibfield  {author} {\bibinfo {author} {\bibfnamefont {F.}~\bibnamefont
  {Hofbauer}},\ }\href@noop {} {\bibinfo {title} {A simple proof of feuerbach's
  theorem}} (\bibinfo {year} {2016}),\ \Eprint
  {https://arxiv.org/abs/1610.03962} {arXiv:1610.03962 [math.HO]} \BibitemShut
  {NoStop}%
\bibitem [{\citenamefont {Hess}\ \emph {et~al.}(2016)\citenamefont {Hess},
  \citenamefont {Sch{\"a}fer},\ and\ \citenamefont {Greiner}}]{hess2016pseudo}%
  \BibitemOpen
  \bibfield  {author} {\bibinfo {author} {\bibfnamefont {P.~O.}\ \bibnamefont
  {Hess}}, \bibinfo {author} {\bibfnamefont {M.}~\bibnamefont {Sch{\"a}fer}},\
  and\ \bibinfo {author} {\bibfnamefont {W.}~\bibnamefont {Greiner}},\
  }\href@noop {} {\emph {\bibinfo {title} {Pseudo-Complex General
  Relativity}}}\ (\bibinfo  {publisher} {Springer},\ \bibinfo {year}
  {2016})\BibitemShut {NoStop}%
\end{thebibliography}%
\end{document}